\documentclass[fleqn,twoside]{article}
\usepackage{espcrc2}

\usepackage{graphicx}

\newcommand{\ssize}[1]{\mbox{\footnotesize $\displaystyle {#1}$}}
\newcommand{\ihsp}{\hspace*{\fill} }

\newcommand{\be}{\begin{equation}}
\newcommand{\ee}{\end{equation}}
\newcommand{\bea}{\begin{eqnarray}}
\newcommand{\eea}{\end{eqnarray}}

\newcommand{\BS}{Bethe--Salpeter }

\newcommand{\Eq}[1]{Eq.~(\ref{#1})}

\newcommand{\Fig}[1]{Fig.~{\ref{#1}}}
\newcommand{\Figs}[1]{Figs.~{\ref{#1}}}
\newcommand{\Ref}[1]{Ref.~\cite{#1}}

\newcommand{\G}{\Gamma}

\newcommand{\ga}{\gamma}
\newcommand{\ro}{\rho}
\newcommand{\la}{\lambda}

\def\slr#1{\setbox0=\hbox{$#1$}           
   \dimen0=\wd0                                 
   \setbox1=\hbox{/} \dimen1=\wd1               
   \ifdim\dimen0>\dimen1                        
      \rlap{\hbox to \dimen0{\hfil/\hfil}}      
      #1                                        
   \else                                        
      \rlap{\hbox to \dimen1{\hfil$#1$\hfil}}   
      /                                         
   \fi}

\title{Nonperturbative QCD Phenomenology and Light Quark Physics}

\author{P.~C.~Tandy\address{Center for Nuclear Research, 
Department of Physics,  Kent State University, Kent OH 44242, USA}%
}
       
\begin{document}

\begin{abstract}
Recent progress in modeling QCD for hadron physics through truncated 
Dyson-Schwinger equations is reviewed.  Special emphasis is put upon 
comparison of dressed quark propagators and the dressed quark-gluon 
vertex with lattice-QCD results. 
\vspace{1pc}
\end{abstract}

\maketitle

\section{Quark DSE solutions and Lattice Data}

The renormalised dressed-quark propagator, $S(p)$, is the solution of the
Dyson-Schwinger equation (DSE)
\begin{equation} 
S^{-1}(p)\!=\!Z_2 \, i\, \gamma \cdot p + Z_4\, m(\mu) 
 + \Sigma'(p,\Lambda), 
    \label{quarkdse} 
\end{equation} 
wherein the dressed-quark self-energy is$^{\,1}$\footnotetext[1]{We use a
Euclidean metric where the scalar product of two four vectors is
\mbox{$a\cdot b=\sum_{i=1}^4 a_i b_i$}, and Hermitian Dirac-$\gamma$ matrices
that obey \mbox{$\{\gamma_\mu,\gamma_\nu\} = 2\delta_{\mu\nu}$}.}
\begin{equation} 
\Sigma'(p,\Lambda) = Z_{\rm 1F} \!  \int^\Lambda_q \!\!\!
 g^2D_{\mu\nu}(k) \, \frac{\lambda^i}{2}\gamma_\mu \, S(q) \,
 \Gamma^i_\nu(q,p).  \label{quarkSelf}
\end{equation} 
Here $D_{\mu\nu}(k)$ is the renormalised dressed gluon propagator;
\mbox{$k=p-q$}; 
$\Gamma^i_\nu(q,p)$ is the renormalised dressed quark-gluon vertex; and
$Z_{\rm 1F}(\mu,\Lambda)$, $Z_2(\mu,\Lambda)$ and $Z_4(\mu,\Lambda)$ are,
respectively, renormalisation constants for the quark-gluon
vertex, quark wave function and current-quark mass.
In \Eq{quarkSelf}, \mbox{$\int^\Lambda_q :=$} \mbox{$\int^\Lambda d^4
q/(2\pi)^4$} denotes an integral with a translationally invariant 
ultraviolet regularization at mass-scale $\Lambda$.  
 
The general form of the dressed-quark propagator is  
\mbox{$S(p) =$} \mbox{$Z(p^2;\mu^2)/[i \gamma \cdot p  +  M(p^2)]$}
where $Z(p^2;\mu^2)$ and $M(p^2)$ are the wave function 
renormalization function and running mass-function respectively.  
The $Z_2$ and $Z_4$ are determined by solving the gap equation, Eq.\
(\ref{quarkdse}), subject to the renormalization condition 
\mbox{$S^{-1}(p)\big|_{p^2=\mu^2} =$} \mbox{$i \gamma \cdot p + m(\mu)$} 
at some large spacelike $\mu^2$.  

The gap equation, Eq.\ (\ref{quarkdse}), is coupled to the DSEs
satisfied by the gluon 2-point function and the vertex 3-point function
which in turn  involve other $n$-point functions; a tractable model 
is defined by a truncation.  At least one nonperturbative, chiral
symmetry preserving truncation exists \cite{Bender:1996bb,Bender:2002as} and the
first term in that scheme is the renormalization-group-improved rainbow gap
equation, wherein the self-energy, Eq.\ (\ref{quarkSelf}), assumes the form
\begin{equation} 
\int_q^\Lambda \! {\cal G}(k^2)\, D_{\mu\nu}^{\rm free} (k) 
\frac{\lambda^a}{2}\gamma_\mu \, S(q) \, \frac{\lambda^a}{2} \gamma_\nu\,, 
\label{DSEAnsatz} 
\end{equation} 
where $k= p-q$ and $D_{\mu\nu}^{\rm free} (k)$ is the Landau gauge free-gluon
propagator.

In Eq.\ (\ref{DSEAnsatz}), ${\cal G}(k^2)$ is an effective coupling, onto
which has been mapped the combined effect of dressing both the gluon propagator 
and quark-gluon vertex.    In the ultraviolet, ${\cal G}(k^2)$ matches 
$4 \pi \alpha_s^{\rm 1-loop}(k^2)$.  It has been found that a
one-parameter infrared form for ${\cal G}(k^2)$ can produce an excellent
description for the ground state pseudoscalar and vector mesons and their
electroweak form factors and decays~\cite{Maris:1999nt,Maris:2003vk}.    
The advent of lattice-QCD data for the quark~\cite{Bowman:2002kn} and 
gluon~\cite{Leinweber:1998uu} propagators provides an opportunity for an
examination of the QCD content of ${\cal G}(k^2)$.
\begin{figure*}[ht]
\centering{\
\includegraphics[width=0.8\textwidth,angle=0]{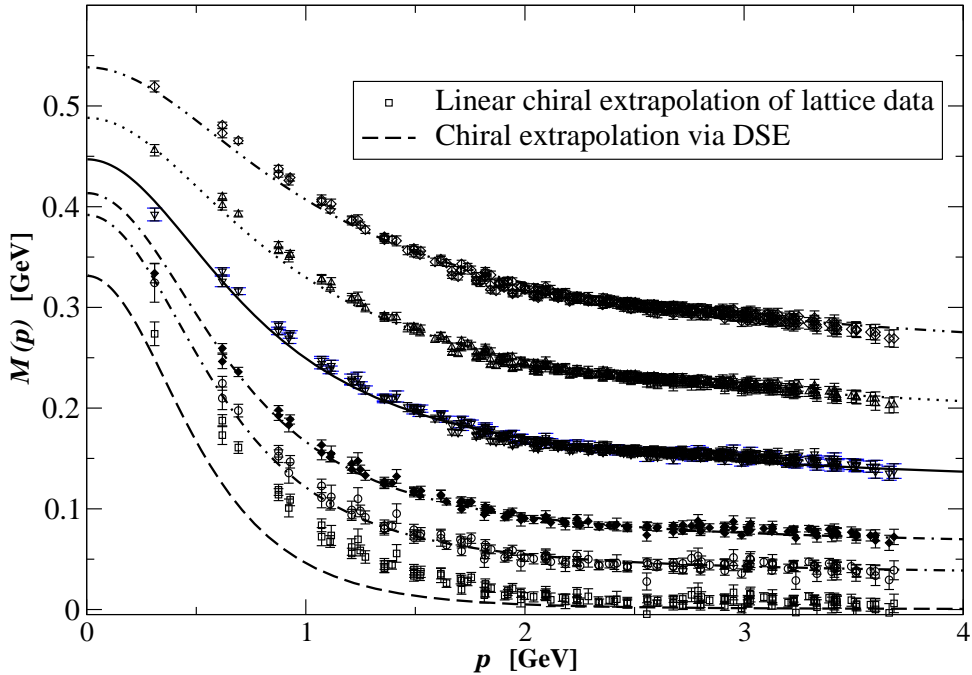}
}
\vspace*{-5mm}
\caption{DSE solutions with vertex dressing fit to lattice data for the 
quark mass function for various current quark masses. \label{fig:Mfnfits} }
\end{figure*}

A first effort has been made to unravel the infrared contribution to 
${\cal G}(k^2)$ from the dressed quark-gluon vertex~\cite{Bhagwat:2003vw}. 
Since the lattice gluon propagator is infrared-suppressed, the required
vertex will be infrared-enhanced to be empirically successful.
In that work, an effective kernel ${\cal G}(k^2)$ is constructed from 
the quenched lattice~\cite{Leinweber:1998uu} Landau gauge gluon propagator 
$D^{\rm latt}(k^2)$ and an effective infrared behavior of the vertex  
determined by requiring that the DSE reproduce the quenched lattice 
data~\cite{Skullerud:2003qu} for the quark propagator.     The kernel is
given the form  
\begin{equation} 
\label{effective} 
\frac{1}{k^2}\, {\cal G}(k^2) =  D^{\rm latt}(k^2)\, \Gamma_1(k^2)\,, 
\end{equation} 
where the dressed vertex is represented by $\gamma_\nu\,\Gamma_1(k^2)$ with
\begin{equation} 
\label{vQ2} \Gamma_1(k^2)= \frac{4 \pi^2\gamma_m}{Z_g}\,\frac{
[\frac{1}{2}\ln(\tau + k^2/\Lambda_{g}^2)]^{d_D}} {[\ln(\tau +
k^2/\Lambda_{\rm QCD}^2)]}\, v(k^2)\,,
\end{equation} 
and \mbox{$\tau=e^2-1>1$}.   The phenomenological factor $v(k^2)$ is unity in the
ultraviolet limit and the log factor combines with the leading log behavior 
of the gluon propagator to ensure
that the entire kernel matches the 1-loop QCD behavior of the running
coupling  \mbox{${\cal G}(k^2) \to$} \mbox{$4\pi^2\gamma_m/\ln( k^2/\Lambda_{\rm 
QCD}^2)$}.  For the quenched, Landau-gauge study, $N_f=0$, \mbox{$d_D= 13/22$} 
and \mbox{$\gamma_{m}=$} \mbox{$12/(33 - 2 N_{f})$} is the anomalous mass 
dimension.  
\begin{figure}[ht] 
\centerline{\resizebox{0.45\textwidth}{!}{\includegraphics{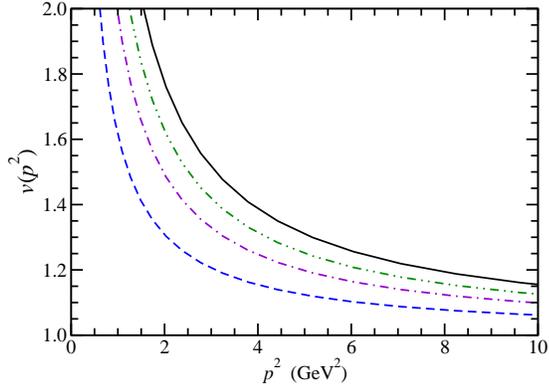}}}
\vspace*{-5mm} 
\caption{\label{Vq2pic} Vertex factor obtained in the 
chiral limit (\textit{solid curve}) and with the three lowest current-quark 
masses from Eq.\ (\protect\ref{amvalues}).  $v(0)$ is finite. }
\vspace*{-5mm}
\end{figure} 

Studies of corrections to the rainbow
truncation show $\Gamma_1$ to be the dominant amplitude of the dressed
vertex in respect to strength and effect on observables. 
It is the only amplitude that is ultraviolet divergent at one-loop level. 
The current-quark mass, $m(\mu)$, was fixed by matching the DSE and
lattice results~\cite{Bowman:2002kn} for $M(p^2)$ on $p^2 > 1\,$GeV$^2$ for
each lattice set characterized by $ma$.  With a DSE renormalisation scale 
$\mu= 19\,$GeV and with $m(\mu)$ in GeV, we find
\begin{equation} 
\label{amvalues}
\begin{array}{l|lllll} 
a\,m     & 0.018 & 0.036 & 0.072 & 0.108 & 0.144 \\\hline 
m(\mu) & 0.030 & 0.055 & 0.110 & 0.168 & 0.225 
\end{array}\,. 
\end{equation}

In Fig.\ \ref{fig:Mfnfits} we compare DSE solutions for $M(p^2)$ with lattice 
results.  It is apparent that the lattice-gluon and lattice-quark propagators 
are easily correlated via a simple infrared vertex enhancement $v(k^2)$ 
depicted in 
Fig.\ \ref{Vq2pic}.   Such vertex enhancement has been anticipated 
for some time based on studies of the connection between DCSB and 
observables~\cite{Hawes:1994ef,Hawes:1998cw}.   Note that $v(0)$ is finite;
for the $m(\mu)$ case considered in Sec.~3 in relation to lattice data for 
the vertex, \mbox{$v(k^2_{\rm s})=$} \mbox{$13.0$}, where 
\mbox{$k^2_{\rm s} \sim 0.04~{\rm GeV}^2$} is an effective infrared scale.
\begin{figure}[ht] 
\centerline{\resizebox{0.45\textwidth}{!}{\includegraphics{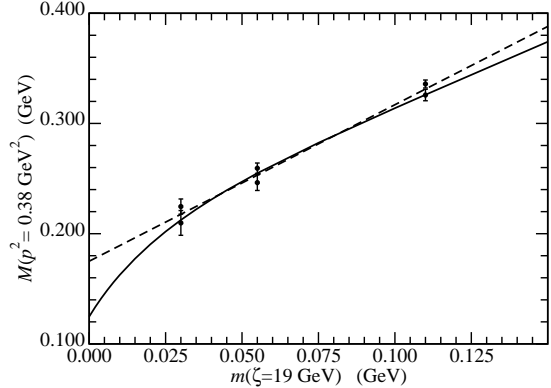}}} 
\vspace*{-5mm}
\caption{Chiral extrapolation.  Solid curve, the lattice-assisted DSE result; 
circles, lowest mass lattice data~\protect\cite{Bowman:2002kn}; dashed-line, 
linear fit to the lattice data.
\label{Fig:Mvsm} } 
\vspace*{-5mm}
\end{figure} 

With the dependence of the lattice data upon $p$ and $m(\mu)$ mapped into 
a DSE kernel, it is a vehicle for chiral extrapolaton.    
It is apparent from \Fig{fig:Mfnfits}  that the linear 
extrapolation~\cite{Bowman:2002bm} overestimates the chiral DSE mass function.
We make a finer comparison in Fig.\ \ref{Fig:Mvsm} which displays 
$M(p^2=p_{\rm IR}^2)$ versus $m(\mu)$ 
where $p^2_{\rm IR} = 0.38\,{\rm GeV}^2$, is the smallest $p^2$ value  
containing two lattice results for $M(p^2)$ at the three lowest $m(\mu)$ values.
The linear trajectory
suggested by the lattice results provides a poor extrapolation to 
$m(\mu) = 0$, giving a result $40$\% too 
large$^{\rm\footnotemark[2]}$\footnotetext[2]{For an analysis in terms of the
DSE chiral susceptibility see \Ref{Holl:2004qn}.}.
The linearly-extrapolated lattice data produces the 
estimate~\cite{Bowman:2002kn}:
$-\langle\bar{q}q\rangle^0_{1\,{\rm GeV}}= (0.270\pm 0.027\,{\rm GeV})^3$.
However, the error is purely statistical.  The systematic error, to which the
linear extrapolation must contribute, was not estimated.  
The DSE-assisted chiral limit~\cite{Bhagwat:2003vw} indicates that  in 
quenched-QCD $-\langle\bar{q}q\rangle^0_{1\,{\rm GeV}}=
(0.19\,{\rm GeV})^3$, a factor of two smaller than the empirical value 
$(0.24 \pm 0.01\,{\rm GeV})^3$.  Related to this,
the quenched lattice data underestimates the chiral  
$f_\pi^0$ by 30\%~\cite{Bhagwat:2003vw}. 

\section{Beyond Ladder-rainbow}

Since the pseudoscalar states are dominated by chiral symmetry and its pattern of 
dynamical and explicit breakdown, most work on extensions of the BSE kernel
beyond the rainbow-ladder level have sought to preserve this symmetry.
Particular extensions of the self-energy beyond rainbow must be accompanied
by related extensions of the ladder kernel so that the axial vector
Ward-Takahashi is preserved.   In the presence of an empirical amount of DCSB,
this identity ensures the Goldstone nature of the chiral limit 
pseudoscalars, and the physical mass patterns,  independently of model 
details~\cite{Maris:1998hd}.  This may be implemented constructively if a 
model of the dressed quark-gluon vertex is defined in terms of Feynman diagrams. 
Several studies~\cite{Bender:1996bb,Bender:2002as,Bhagwat:2004hn} have 
employed the Abelian-like series 
\bea
&&Z_{\rm 1F}\, \frac{\la^a}{2}\G_{\nu}(k,p) = 
\frac{\la^a}{2}\ga_{\nu} - \int_q^\Lambda\, D_{\ro\la}(q)\, 
\frac{\la^b}{2}\ga_{\ro}\,\nonumber \\
&&\times S(k+q) \,\frac{\la^a}{2}\ga_{\nu}\, S(p+q) \,\frac{\la^b}{2}
\ga_{\la} + \cdots~~,
\label{eq:qgv0}
\eea
which has the ladder-rainbow truncation of \Eq{quarkSelf} as the first term.
\begin{figure}[ht]
\vspace*{-5mm}
\centering{\
\includegraphics[width=0.45\textwidth]{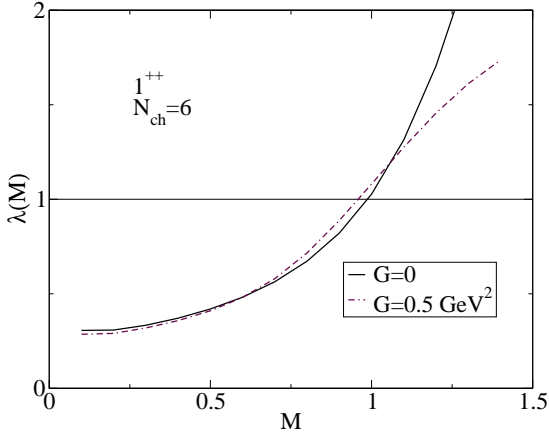} }
\vspace*{-9mm}
\caption{\label{fig:axp}  $1^{++}$ ($a_1/f_1$) axialvector  meson BSE 
eigenvalue from the  ladder kernel (\mbox{$G=0$}) and from the dressed 
vertex model (\mbox{$G > 0$}). }
\vspace*{-7mm}
\end{figure}

The corresponding \BS kernel in this scheme is generated  as the sum of 
terms produced by cutting a quark line in the resulting diagrammatic 
representation of the quark self-energy.
Due to the complexity of a BSE calculation with a multi-loop
kernel, most numerical 
implementations~\cite{Bender:1996bb,Bender:2002as,Bhagwat:2004hn} have 
exploited the algebraic  structure that follows from use of the infrared 
dominant model \mbox{$D_{\mu \nu}(q) \propto $} 
\mbox{$G T_{\mu \nu}(q) \delta^4(q)$}.   For infrared-dominated physics this 
model effectively summarizes the behavior of
more realistic models.   A disadvantage is that it does not support finite
quark-quark relative momentum in bound states.  This is likely the reason 
why it does not bind the (P-wave) axial-vector states.

Taking advantage of the fact that existing finite range ladder-rainbow 
models typically produce axial vector masses of about 0.9~GeV, 
Ref.~\cite{Watson:2004kd} has used the MN model only as an effective 
representation
of the gluon exchange that implements vertex dressing via \Eq{eq:qgv0}.  
The residual effective gluon exchange is represented by an existing
finite range model~\cite{Alkofer:2002bp}.
With a simple adjustment, the chiral symmetry-preserving BSE kernel is
easily obtained.   This does not implement the ultraviolet behavior
of the QCD running coupling; it contributes typically  10\% or less to 
meson masses~\cite{Maris:1999nt} and this level of precision is not
our concern here.   
\begin{figure}[ht]
\vspace*{-5mm}
\centering{\
\includegraphics[width=0.45\textwidth]{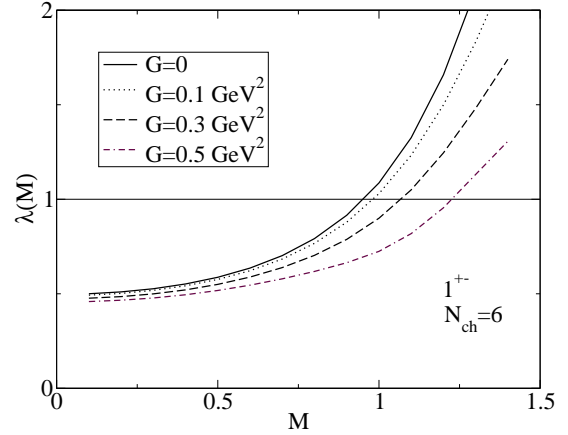} }
\vspace*{-9mm}
\caption{\label{fig:axm}  $1^{+-}$ ($b_1/h_1$) axialvector  meson BSE 
eigenvalue from the ladder kernel (\mbox{$G=0$}) and from the dressed 
vertex model (\mbox{$G > 0$}).}
\vspace*{-7mm}
\end{figure}

The eigenvalue behavior for the axial vector solutions in  the 
\mbox{$1^{++}$} ($a_1/f_1$) and 
\mbox{$1^{+-}$} ($b_1/h_1$) channels are displayed in \Figs{fig:axp} 
and \ref{fig:axm}.  In previous work, 
the ladder truncation, constrained by 
chiral data, is generally found to be 200-400~MeV too attractive for 
these P-wave states~\cite{MarisPrivCom,Jarecke:2002xd,Alkofer:2002bp}. 
Our present results agree with this.
The \mbox{$1^{++}$} channel shows a 30~MeV of attraction
due to the effect of 1-loop dressing added to the ladder kernel.   However,
in the \mbox{$1^{+-}$} ($b_1/h_1$) channel, we find a repulsive effect
of 290~MeV above the ladder kernel result, yielding a value close
to $m^{\rm expt}_{{\rm b}_1}$.   We are not able to compare 
these findings with previous work on vertex dressing since such P-wave 
states do not have solutions in the models considered previously for
that purpose~\cite{Bender:1996bb,Bender:2002as,Bhagwat:2004hn}.
Other studies of $a_1$ and $b_1$ based on the ladder-rainbow truncation
have used a separable approximation where the quark propagators are the 
phenomenological instruments~\cite{Burden:1997nh,Bloch:1999vk,Burden:2002ps}, 
these studies find more acceptable masses for both states in the vicinity of 
1.3~GeV.

\section{Nonperturbative Model of the Quark-gluon Vertex}

\begin{figure}[ht]
\vspace*{-20mm}
\centering{\ \hspace*{-5mm}
\includegraphics[width=40mm]{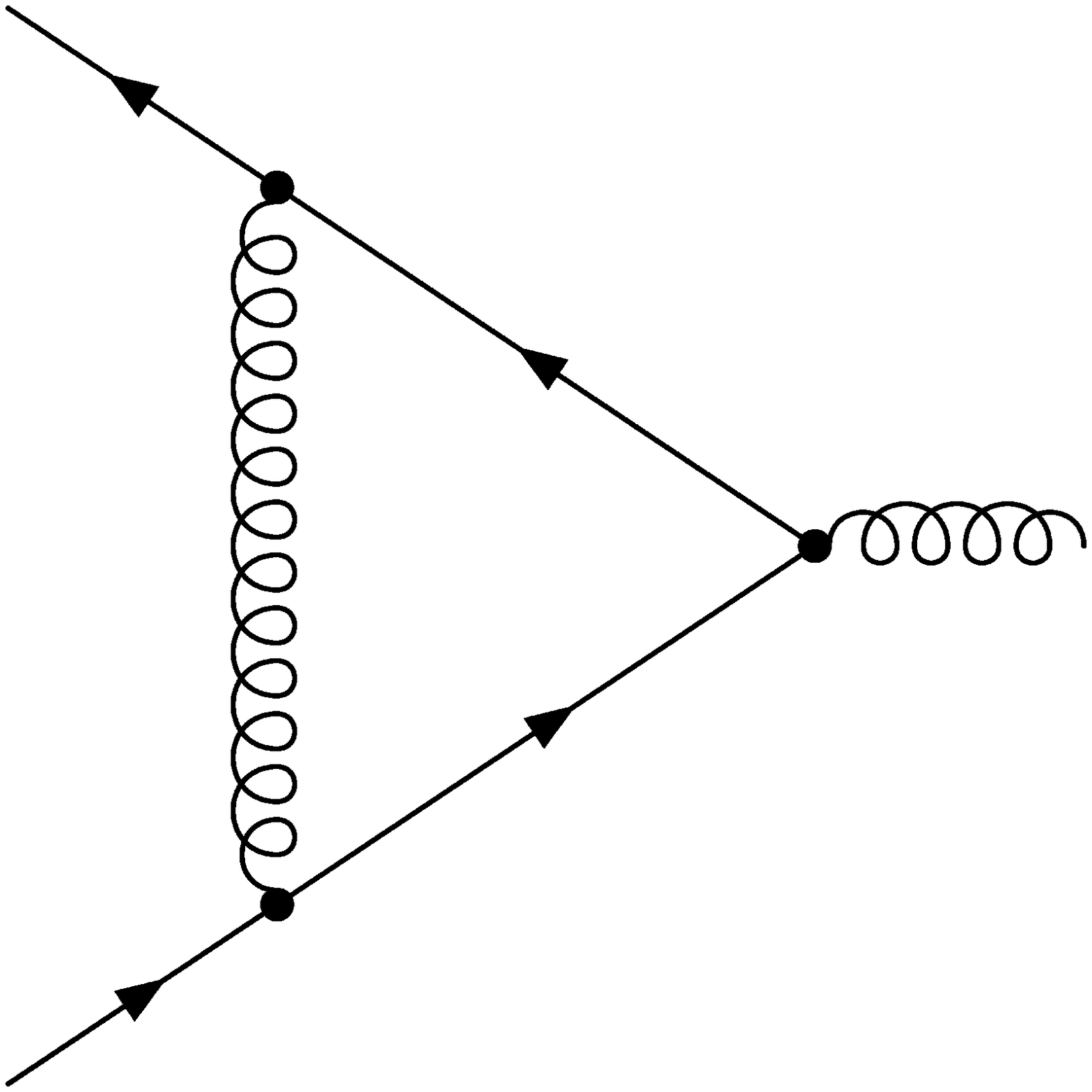}\hspace*{-8mm}
\includegraphics[width=40mm]{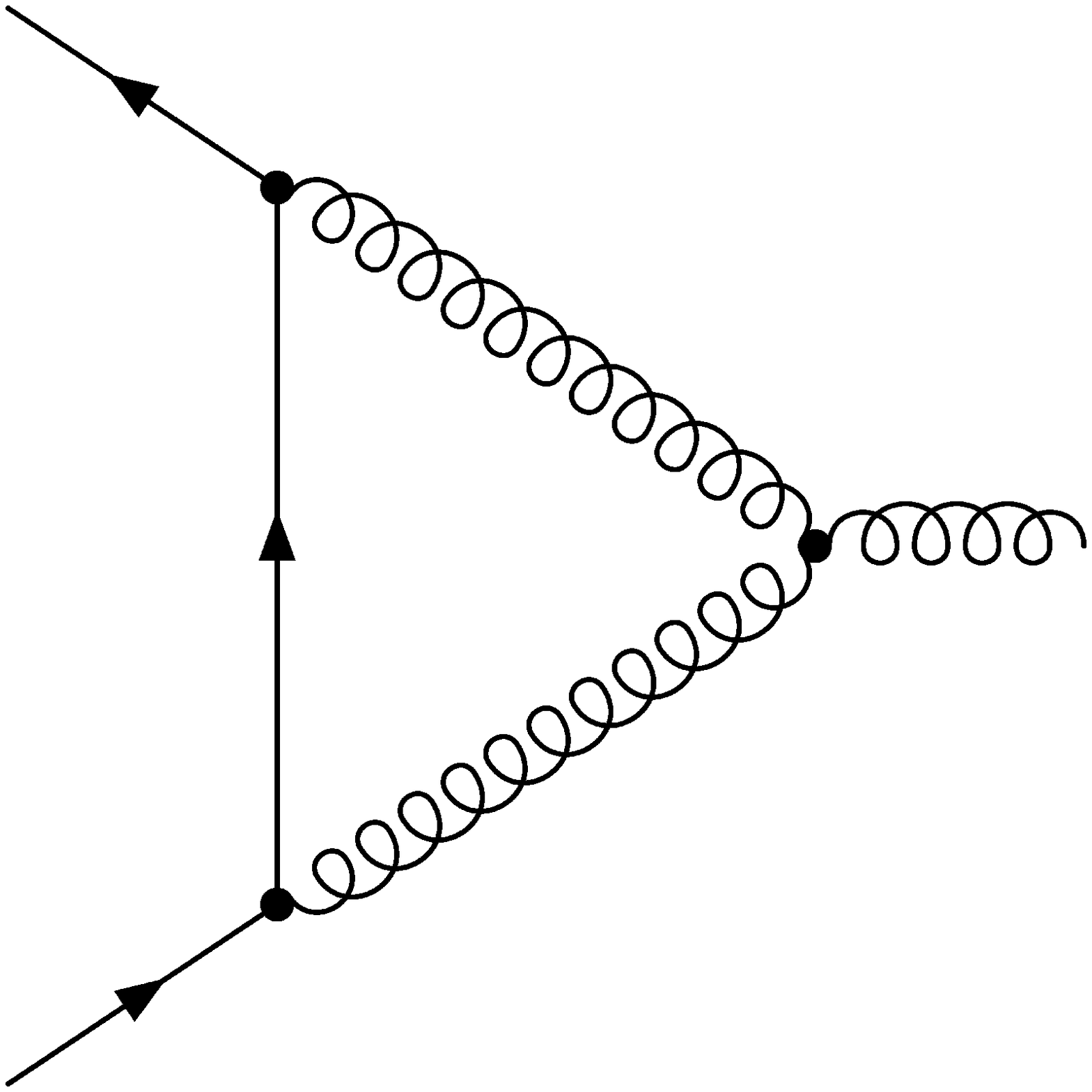} }
\vspace*{-20mm}
\caption{\label{fig:2vertdiags} The dressing of the quark-gluon vertex at 
one loop: the Abelian-like term $\Gamma_\sigma^{\rm A}$, and the non-Abelian 
term $\Gamma_\sigma^{\rm NA}$. }
\vspace*{-20mm}
\centerline{\ihsp A \ihsp \hspace*{20mm} NA \ihsp}
\vspace*{10mm}
\end{figure}

We denote the  dressed-quark-gluon vertex for gluon momentum $k$ and quark 
momentum $p$ by \mbox{$ig\, t^c\,\Gamma_\sigma(p+k,p)$}, where 
\mbox{$t^c = \lambda^c/2$} and $\lambda^c$ is an SU(3) color matrix.  
Through ${\cal O}(g^2)$, i.e., to 1-loop, the amplitude $\Gamma_\sigma$ is given,
in terms of \Fig{fig:2vertdiags}, by 
\mbox{$\Gamma_\sigma(p+k,p)  = $} \mbox{$ Z_{\rm 1F}\,\gamma_\sigma + 
\Gamma_\sigma^{\rm A} + \Gamma_\sigma^{\rm NA} + \ldots$}.

The color factors reveal two important considerations.  The color factor of the
(Abelian-like) term $\Gamma_\sigma^{\rm A}$ would be given by
\mbox{$t^a\,t^a = \,\ssize{C_{\rm F}} = \ssize{(N_c^2-1)}/ \ssize{2 N_c}$}
for the strong dressing of the photon-quark vertex, i.e., in the color
singlet channel.  The octet $\Gamma_\sigma^{\rm A}$ is of opposite sign and is 
suppressed 
by a factor $1/(N_c^2-1)$: single gluon exchange between a quark and antiquark 
has relatively weak repulsion in the color-octet channel, compared to strong
attraction in the color-singlet channel.  Net attraction for the gluon vertex
(at least to this order)  is provided by the non-Abelian 
$\Gamma_\sigma^{\rm NA}$ term, which involves the three-gluon vertex:
the color factor is amplified by $-N_c^2$ over the $\Gamma_\sigma^{\rm A}$
term.

\begin{figure}[ht] 
\vspace*{-8mm}
\centerline{\includegraphics[width=0.45\textwidth]{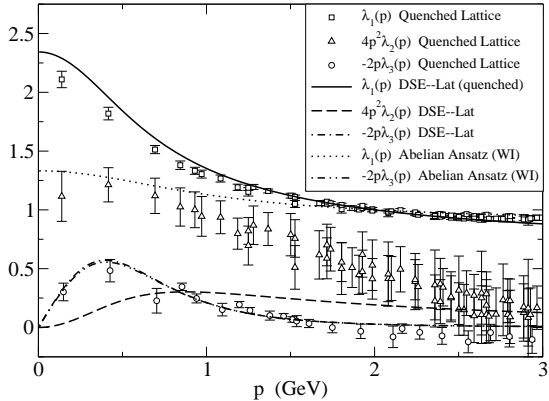}}
\vspace*{-5mm}
\caption{\label{fig:Lat_WI} The quark-gluon vertex amplitudes at zero 
gluon momentum and for quark current mass 
\mbox{$m(\mu=2~{\rm GeV}) =$} 60~MeV.  Quenched lattice 
data~\protect\cite{Skullerud:2003qu} is compared to the results of the
DSE-Lat model~\protect\cite{Bhagwat:2003vw}.  }
\vspace*{-5mm}
\end{figure} 

At \mbox{$k=0$} we note that the bare result 
\mbox{$\Gamma^{3g}_{\mu \nu \sigma}(q,q) =$} 
\mbox{$-\partial [D_0^{-1}(q) T_{\mu \nu}(q)]/\partial q_\sigma$}
allows the total gluon content of the integrand of $\Gamma_\sigma^{\rm NA}$ 
to be 
\mbox{$[\partial g^2\,D_0(q^2)/\partial q_\sigma] T_{\mu \nu}(q)$}.
The dressing provided by the combination 
\mbox{$\Gamma_\sigma^{\rm A} + \Gamma_\sigma^{\rm NA}$}  
yields a vertex that satisfies the Slavnov-Taylor identity (STI) through 
${\cal O}(g^2)$~\cite{Davydychev:2000rt}.  

Our nonperturbative model~\cite{Bhagwat:2004kj} for the dressed quark-gluon 
vertex is defined by extentions of the two diagrams 
\mbox{$\Gamma_\sigma^{\rm A} + \Gamma_\sigma^{\rm NA}$}  into 
dressed versions 
determined solely from an existing ladder-rainbow model DSE kernel that has
1-loop QCD renormalization group improvement.  Two DSE models are 
employed.  The first (DSE-Lat)~\cite{Bhagwat:2003vw} is the
mapping of quenched lattice data for the  gluon 
propagator into a continuum ladder-rainbow kernel with effective gluon vertex
as described in Sec.~1.   In this sense, it represents quenched dynamics.  
The second 
(DSE-MT)~\cite{Maris:1999nt} provides a good one-parameter fit to a wide 
variety of light quark meson physics; in this sense it represents unquenched 
dynamics.    Both are implemented through the
substitution \mbox{$g^2\, D_0(q^2) \to$} \mbox{${\cal G}(q^2)/q^2$}
with bare quark propagators  replaced
by rainbow DSE solutions.   Note that  this 
procedure makes a corresponding nonperturbative extension of
$\Gamma^{3g}_{\mu \nu \sigma}$.   Our justification is one of simplicity;  no 
new parameters are introduced.

The general nonperturbative vertex at \mbox{$k=0$} has a representation in 
terms of three invariant amplitudes;  
\mbox{$\Gamma_\sigma(p,p) =$} \mbox{$ \gamma_\mu \lambda_1(p^2)
- 4 p_\mu\, \gamma \cdot p\, \lambda_2(p^2) - i 2 p_\mu\, \lambda_3(p^2)$};
the lattice-QCD data~\cite{Skullerud:2003qu} is provided in terms of 
the $\lambda_i(p^2)$.   A useful comparison is the Abelian  Ward identity 
\mbox{$\Gamma_\sigma^{WI}(p,p) =$} 
\mbox{$ -i \partial S^{-1}(p)/\partial p_\sigma$}.  

In \Fig{fig:Lat_WI} we display the DSE-Lat model results in a
dimensionless form for comparison with the (quenched) lattice 
data$^{\rm\footnotemark[3]}$\footnotetext[3]{We note 
that in \Ref{Skullerud:2003qu} both the lattice data, and the Abelian 
(Ward identity) Ansatz, for $\lambda_3(p)$ are presented as positive.  
These two sign errors have been acknowledged~\cite{Skull_PrivCom04}.}.  
The renormalization scale of the lattice data is \mbox{$\mu = 2$}~GeV
where \mbox{$\lambda_1(\mu) = 1$}, \mbox{$A(\mu) = 1$}.   We compare to the 
lattice data set for which \mbox{$m(\mu) = 60$}~MeV.  The same
renormalization scale and conditions have been implemented in the DSE 
calculations.   Without parameter adjustment, the model reproduces the
lattice data for $\lambda_1$ and $\lambda_3$ quite well.   The Abelian Ansatz 
(Ward Identity), while clearly inadequate for $\lambda_1$ below 1.5~GeV,
reproduces $\lambda_3$.    The DSE model $\lambda_2$ is evidently too
weak, although the lattice data has large errors.    The evident infrared
enhancement at \mbox{$k^2_{\rm s} \sim 0.04~{\rm GeV}^2$} is 
\mbox{$\lambda_1(k^2_{\rm s}) \sim 2.3$}, a factor of 6 smaller than what
the analysis of the quark propagator in Sec.~1 required.   The
non-Abelian term $\Gamma^{\rm NA}_\sigma$ dominates to a greater extent 
than what the ratio of color factors ($-9$) would suggest; it also 
distributes its infrared strength to favor $\lambda_1$ more so than does
$\Gamma^{\rm A}_\sigma$.    The present model results could be summarized
quite effectively by ignoring $\Gamma^{\rm A}_\sigma$.
Due to the definition of the two DSE models, their 
comparison~\cite{Bhagwat:2004kj}   provides an estimate of the effects 
of the quenched approximation.  The effects are comparable to present 
uncertainties in the lattice data.
 
Recent work on a model of the quark-gluon vertex using a bare
$\Gamma^{3g}_{\mu \nu \sigma}$ vertex with DSE solutions for the gluon 
propagator and an Ansatz for dressing both internal quark-gluon vertices 
has produced results similar to the present work, except that the 
\mbox{$m(\mu) = 115$}~MeV case is considered~\cite{Fischer:2004ym}.
Evidently the detailed infrared structure of $\Gamma^{3g}_{\mu \nu \sigma}$ is 
not crucial to present considerations.

{\bf Acknowledgements}:~~We thank R.~Alkofer, C.~S.~Fischer, and C.~D.~Roberts 
for very useful discussions and are grateful to P.~Bowman and J.~I.~Skullerud 
for providing the lattice-QCD results.   Appreciation is extended to  
A.~{K\i z\i lers\"u} and the organizing committee for making this workshop 
possible, and to the staff and members of the CSSM, Adelaide for hospitality 
and support.  This work has 
been partially supported by  NSF grants no. PHY-0301190 and no. INT-0129236.  


\bibliographystyle{apsrev}

\bibliography{refsPM,refsPCT,refsCDR,refs}

\end{document}